\RequirePackage{ifpdf}
\ifpdf 
\documentclass[pdftex]{sigma}
\else
\documentclass{sigma}
\fi

\begin{document}
\allowdisplaybreaks

\renewcommand{\PaperNumber}{083}

\FirstPageHeading

\renewcommand{\thefootnote}{$\star$}

\ShortArticleName{Fermion on Curved Spaces, Symmetries, and
Quantum Anomalies}

\ArticleName{Fermion on Curved Spaces, Symmetries,\\
and Quantum Anomalies\footnote{This paper is a contribution to the
Proceedings of the O'Raifeartaigh Symposium on Non-Perturbative
and Symmetry Methods in Field Theory (June 22--24, 2006, Budapest,
Hungary). The full collection is available at
\href{http://www.emis.de/journals/SIGMA/LOR2006.html}{http://www.emis.de/journals/SIGMA/LOR2006.html}}}

\Author{Mihai VISINESCU}

\AuthorNameForHeading{M. Visinescu}

\Address{Department of Theoretical Physics, Institute for Physics
and Nuclear Engineering, Magurele, P.O.Box MG-6, Bucharest,
Romania}
\Email{\href{mailto:mvisin@theory.nipne.ro}{mvisin@theory.nipne.ro}}
\URLaddress{\href{http://www.theory.nipne.ro/~mvisin/}{http://www.theory.nipne.ro/\~{}mvisin/}}

\ArticleDates{Received September 28, 2006, in f\/inal form
November 21, 2006; Published online November 29, 2006}

\Abstract{We review the geodesic motion of pseudo-classical
spinning particles in curved spaces. Investigating the generalized
Killing equations for spinning spaces, we express the constants of
motion in terms of Killing--Yano tensors. Passing from the
spinning spaces to the Dirac equation in curved backgrounds we
point out the role of the Killing--Yano tensors in the
construction of the Dirac-type operators. The general results are
applied to the case of the four-dimensional Euclidean
Taub--Newman--Unti--Tamburino space. The gravitational and axial
anomalies are studied for generalized Euclidean Taub-NUT metrics
which admit hidden symmetries analogous to the Runge--Lenz vector
of the  Kepler-type problem. Using the Atiyah--Patodi--Singer
index theorem for manifolds with boundaries, it is shown that the
these metrics make no contribution to the axial anomaly.}


\Classification{83C47; 83C40; 83C20}

\section{Introduction}

The aim of this paper is to investigate the quantum objects,
namely spin one half particles, in curved spaces. Having in mind
the lack of a satisfactory quantum theory for gravitational
interaction, this study is justif\/ied and not at all trivial.

In order to study the geodesic motions and the conserved classical
and quantum quantities for fermions on curved spaces, the
symmetries of the backgrounds proved to be very important. We
mention that the following two generalization of the Killing (K)
vector equation have become of interest in physics:
\begin{enumerate}\itemsep=0pt
\item A symmetric tensor f\/ield $K_{\mu_1\dots\mu_r}$  is called
a St\" ackel--Killing (S-K) tensor of valence $r$ if and only if
\begin{equation}\label{SK}
K_{(\mu_1\dots\mu_r;\lambda)} = 0.
\end{equation}
The usual Killing (K) vectors correspond to valence $r=1$ while
the hidden symmetries are encapsulated in S-K tensors of valence
$r>1$. \item A  tensor $f_{\mu_1\dots \mu_r}$ is called a
Killing--Yano (K-Y) \cite{Y} tensor of valence $r$ if it is
totally antisymmetric and it satisf\/ies the equation
\begin{equation}\label{ky}
f_{\mu_1\dots(\mu_r;\lambda)} = 0.
\end{equation}
\end{enumerate}
\looseness=1 These objects can be characterized in several
equivalent ways. For example, K-Y tensors can be def\/ined as
dif\/ferential forms on a manifold whose covariant derivative is
totally antisym\-metric.

There are many important occurrences of K-Y tensors in physics.
For example, investigating the geodesic equations on curved
spaces, the K-Y tensors play an important role in the existence of
the constants of motion \cite{Carter}. K-Y tensors do also appear
in the study of the Klein--Gordon and Dirac equations. There is a
natural and profound connexion between K-Y tensors, Dirac-type
operators and supersymmetries \cite{GRH,a2,Car,a9} having in mind
their anticommuting property.

The models of relativistic particles with spin have been proposed
for a long time and the literature on the particle with spin grew
vast \cite{BeMa,a2,a5}. The models involving only conventional
coordinates are called classical models while the models involving
anticommuting coordinates are generally called pseudo-classical.

In the beginning of this paper we discuss the relativistic spin
one half particle models involving anticommuting vectorial degrees
of freedom which are usually called the spinning particles.
Spinning particles are in some sense the classical limit of the
Dirac particles. The action of spin one half relativistic particle
with spinning degrees of freedom described by Grassmannian (odd)
variables was f\/irst proposed by Berezin and Marinov \cite{BeMa}.

The generalized Killing equations for the conf\/iguration space of
spinning particles (spinning space) are analyzed and the solutions
are expressed in terms of K-Y tensors. We mention that the
existence of a K-Y tensor is both a necessary and a suf\/f\/icient
condition for the existence of a new supersymmetry for the
spinning space \cite{GRH,a3}.

Passing from the pseudo-classical approach to the Dirac equation
in cur\-ved spaces, we point out the role of the K-Y tensors in
the construction of non-standard Dirac-type operators. The
Dirac-type operators constructed with the aid of covariantly
constant K-Y tensors are equivalent with the standard Dirac
operator. The non-covariantly constant K-Y tensors generates
non-standard Dirac operators which are not equivalent to the
standard Dirac operator and they are associated with the hidden
symmetries of the space.

The general results are applied to the case of the
four-dimensional Euclidean Taub--Newman--Unti--Tamburino
(Taub-NUT) space. The Taub-NUT metrics were found by Taub
\cite{Taub} and extended by Newman--Unti--Tamburino \cite{NUT}.
The Euclidean Taub-NUT metric has lately attracted much attention
in physics. Hawking \cite{SH} has suggested  that the Euclidean
Taub-NUT metric might give rise to the gravitational analog of the
Yang--Mills instanton. This metric is the space part of the line
element of the celebrated Kaluza--Klein monopole of Gross and
Perry \cite{GP} and Sorkin \cite{So}. On the other hand, in the
long distance limit, neglecting radiation, the relative motion of
two monopoles is described by the geodesics of this space
\cite{AH}. The Taub-NUT family of metrics is also involved in many
other modern studies in physics like strings, membranes, etc.

The family of Taub-NUT metrics with their plentiful symmetries
provides an excellent background to investigate the classical and
quantum conserved quantities on curved spaces. In the Taub-NUT
geometry there are four K-Y tensors. Three of these are complex
structures realizing the quaternion algebra and the Taub-NUT
manifold is hyper-K\" ahler \cite{GR}. In addition to these three
vector-like K-Y tensors, there is a scalar one which has a
non-vanishing f\/ield strength and which exists by virtue of the
metric being type D. From the symmetry viewpoint, the geodesic
motion in Taub-NUT space admits a ``hidden" symmetry of the Kepler
type.

  The K-Y tensors
play an important role in theories with spin and especially in the
Dirac theory on curved spacetimes where they produce f\/irst order
dif\/ferential operators, called Dirac-type operators, which
anticommute with the standard Dirac one, $D_s$ \cite{CML}. Another
virtue of the K-Y tensors is that they enter as square roots in
the structure of several second rank S-K tensors that generate
conserved quantities in classical mechanics or conserved operators
which commute with $D_s$. The construction of Carter and
McLenaghan depended upon the remarkable fact that the (symmetric)
S-K tensor $K_{\mu\nu}$ involved in the constant of motion
quadratic in the four-momentum $p_\mu$
\begin{equation}\label{i1}
Z = \tfrac{1}{2} K^{\mu\nu} p_\mu p_\nu
\end{equation}
has a certain square root in terms of K-Y tensors $f_{\mu\nu}$:
\begin{equation}\label{i2}
K_{\mu\nu} = f_{\mu\lambda} f^{\lambda~} _{~\nu}.
\end{equation}

These  attributes of the K-Y tensors lead to an ef\/f\/icient
mechanism of supersymmetry especially when the S-K tensor
$K_{\mu\nu}$ in equation (\ref{i1}) is proportional with the
metric tensor $g_{\mu\nu}$ and the corresponding K-Y tensors in
equation (\ref{i2}) are covariantly constant. Then each tensor of
this type, $f^i$, gives rise to a Dirac-type operator,  $D^i$,
representing a supercharge of the superalgebra $\{ D^i , D^j \}
\propto D^{2}_{s} \delta_{ij} $.

The necessary condition that a S-K tensor of valence two be
written as the square of a K-Y tensor is that it has at the most
two distinct eigenvalues \cite{MV}. In the case of the generalized
Taub-NUT spaces \cite{IK1,IK2,IK3} the S-K tensors involved in the
Runge-Lenz vector cannot be expressed as a product of K-Y tensors.
The non-existence of the K-Y tensors on generalized Taub-NUT
metrics leads to gravitational quantum anomalies proportional to a
contraction of the S-K tensor with the Ricci tensor \cite{tanug}.

The index of the Dirac operator is a useful tool to investigate
the topological properties of the manifold as well as in computing
axial quantum anomalies in f\/ield theories. In even-dimensional
spaces one can def\/ine the index of a Dirac operator as the
dif\/ference between the number of linearly independent zero modes
with eigenvalues $+1$ and $-1$ under $\gamma_5$. A remarkable
result states the equality of the indices of the standard and
non-standard Dirac operators \cite{a1}.

For the generalized Taub-NUT spaces, in \cite{tanug} we computed
the axial quantum anomaly, interpreted as the index of the Dirac
operator of these metrics, on annular domains and on disks, with
the non-local Atiyah--Patodi--Singer boundary condition. We also
examined the Dirac operator on the complete Euclidean space with
respect to these metrics, acting in the Hilbert space of
square-integrable spinors. We found  that this operator is not
Fredholm, hence even the existence of a f\/inite index is not
granted.

The structure of the paper is as follows: We start the next
section with a description of the pseudo-classical model for the
spinning particles. In Section 3 we present the spinning Taub-NUT
space. In the next section we consider the Dirac equation on a
curved background with symmetries. In Section 5 the general
results are applied to the case of the Euclidean Taub-NUT space.
In the next two sections we discuss the gravitational and axial
anomalies for generalized Euclidean Taub-NUT metrics. Section 8 is
the conclusion. In two appendices we review the symmetries and
conserved quantities for geodesic motions in the Euclidean
Taub-NUT space and its generalizations.

\section{Pseudo-classical approach}

Spinning space is an extension of an ordinary Riemannian manifold,
pa\-ra\-me\-tri\-zed by local coordinates {$\{$}$x^\mu${$\}$}, to
a graded manifold parametrized by local coordinates {$\{$}$x^\mu,
\psi^\mu${$\}$}, with the f\/irst set of variables being
Grassmann-even (commuting) and the second set Grassmann-odd
(anticommuting) \cite{BeMa}.

The dynamics of spinning point-particles in a curved space-time is
described by the one-dimensional $\sigma$-model with the action:
\[
 S=\int_{a}^{b}d\tau \left(\tfrac{1}{2}g_{\mu \nu}(x)\dot{x}^\mu
\dot{x}^\nu + \tfrac{i}{2} g_{\mu \nu}(x)\psi^\mu
\frac{D\psi^\nu}{ D\tau} \right).
\]

In what follows we shall investigate the conserved quantities for
geodesic motions in the case of spinning manifolds. For this
purpose we consider the world-line Hamiltonian given by
\[
H=\tfrac1 2 g^{\mu\nu}\Pi_\mu \Pi_\nu,
\]
where
\[
\Pi_\mu = g_{\mu\nu}\dot{x}^\nu
\]
is the covariant momentum.

For any constant of motion ${\cal J}(x,\Pi,\psi)$, the bracket
with $H$ vanishes
\[
\left\lbrace H,{\cal J} \right\rbrace = 0,
\]
where the Poisson--Dirac brackets for functions of the covariant
phase space variables $(x,\Pi,\psi)$ is def\/ined by
\[
\left\lbrace F,G\right\rbrace={\cal D}_\mu F\frac{\partial G}
{\partial \Pi_\mu} - \frac{\partial F}{\partial \Pi_\mu}{\cal
D}_\mu G - {\cal R}_{\mu\nu}\frac{\partial F}{\partial
\Pi_\mu}\frac{\partial G}{\partial \Pi_\nu} +
i(-1)^{a_F}\frac{\partial F}{\partial \psi^\mu}\frac{\partial
G}{\partial \psi_\mu}.
\]
The notations used are
\[
{\cal D}_\mu F = \partial_\mu F +
\Gamma^\lambda_{\mu\nu}\Pi_\lambda\frac{\partial F}{\partial
\Pi_\nu} - \Gamma^\lambda_{\mu\nu}\psi^\nu \frac{\partial
F}{\partial \psi^\lambda},\qquad {\cal R}_{\mu\nu} =
\tfrac{i}{2}\psi^\rho\psi^\sigma R_{\rho\sigma\mu\nu}
\]
and $a_F$ is the Grassmann parity of $F$: $a_F=(0,1)$ for
$F=\mbox{(even,odd)}$.

If we expand ${\cal J}(x,\Pi,\psi)$ in a power series in the
covariant momentum
\[
{\cal J}=\sum_{n=0}^{\infty}\frac{1}{n!}{\cal J}^{(n)\mu_1
\dots\mu_n}(x,\psi) \Pi_{\mu_1}\cdots\Pi_{\mu_n}
\]
then the bracket $\{ H , {\cal J}\}$ vanishes for arbitrary
$\Pi_\mu$ if and only if the components of ${\cal J}$ satisfy the
generalized Killing equations \cite{GRH}:
\begin{equation}\label{gke}
{\cal J}^{(n)}_{(\mu_1\dots\mu_n;\mu_{n+1})} + \frac{\partial
{\cal J}^{(n)}_{(\mu_1 \dots\mu_n}}{\partial \psi^\sigma}
\Gamma^\sigma_{\mu_{n+1})\lambda} \psi^\lambda =
\tfrac{i}{2}\psi^\rho \psi^\sigma R_{\rho\sigma\nu(\mu_{n+1}}
{{\cal J}^{(n+1)\nu}}_{\mu_1 \dots \mu_n)},
\end{equation}
where the parentheses denote symmetrization with norm one over the
indices enclosed.

Explicit solutions can be constructed starting with the
geometrical symmetries of the space. For each K vector $R_\mu$,
i.e.\ $R_{(\mu ; \nu)} = 0 $ there is a conserved quantity in the
spinning case:
\begin{equation}\label{j}
{\cal J} = \tfrac i2 R_{[\mu;\nu]} \psi^\mu \psi^\nu + R_\mu
\dot{x}^\mu.
\end{equation}

A more involved example is given by a S-K tensor $K_{\mu\nu}$
satisfying equation (\ref{SK}). Let us assume that a
St\"ackel--Killing tensor $ K_{\mu\nu}$ can be written as a
symmetrized product of two  (dif\/ferent) K-Y tensors $(i\ne j)$
of valence 2:
\[
 K^{\mu\nu}_{ij} = \tfrac{1}{2}(f^\mu_{i~\lambda} f^{\nu\lambda}_j +
f^\nu_{i~\lambda} f^{\mu\lambda}_j) .
\]

The conserved quantity  for the spinning space is \cite{GRH}
\begin{equation}\label{jj}
{\cal J}_{ij} = \tfrac{1}{2!} K^{\mu\nu}_{ij}\dot{x}_\mu
\dot{x}_\nu + {\cal J}^{(1)\mu}_{ij}\dot{x}_\mu + {\cal
J}^{(0)}_{ij},
\end{equation}
where
\begin{gather*}
{\cal J}^{(0)}_{ij} =-\tfrac{1}{4}
\psi^\lambda\psi^\sigma\psi^\rho\psi^\tau
\left(R_{\mu\nu\lambda\sigma} f^\mu_{i~\rho} f^\nu_{j~\tau} +
\tfrac{1}{2} c^{~~~\pi}_{i\lambda\sigma} c_{j\rho\tau\pi}\right)
,
\\
{\cal J}^{(1)\mu}_{ij} = \tfrac{i}{2}\psi^\lambda \psi^\sigma
\left(f^\nu_{i~\sigma} D_\nu f^\mu_{j~\lambda} + f^\nu_{j~\sigma}
D_\nu f^\mu_{i~\lambda} + \tfrac{1}{2} f^{\mu\rho}_i
c_{j\lambda\sigma\rho} + \tfrac{1}{2} f^{\mu\rho}_j
c_{i\lambda\sigma\rho} \right)
\end{gather*}
with
 $c_{i\mu\nu\lambda} = -2 f_{i[\nu\lambda;\mu]}$.

In what follows we shall return to the equation (\ref{gke})
looking for solutions depending exclusively on the Grassmann
variables {$\{$}$\psi^\mu${$\}$}. The existence of such kind of
solutions of the Killing equation is one of the specif\/ic
features of the spinning particle models.

The most remarkable class of solutions is represented by:
\begin{gather}\label{qf}
Q_f = f_{\mu_1 \dots\mu_r}\Pi^{\mu_1}\psi^{\mu_2}\cdots
\psi^{\mu_r} + \frac{i}{r+1}(-1)^{r+1}f_{[\mu_1 \dots
\mu_r;\mu_{r+1}]} \cdot \psi^{\mu_1}\cdots \psi^{\mu_{r+1}}.
\end{gather}

This quantity is a superinvariant
\[
\{ Q_f , Q_ 0 \} = 0,
\]
where $Q_0$ is the supercharge
\begin{equation}\label{Q0}
Q_0=\Pi_\mu\,\psi^\mu .
\end{equation}

Equations (\ref{j}) and (\ref{jj}) describing the conserved
quantities in the spinning space contains specif\/ic spin terms
involving even numbers of Grassmann variables.

\section{Spinning Taub-NUT space}

In the Taub-NUT case, the pseudo-classical approach sets the spin
contributions to the angular momentum, ``relative electric
charge'' (\ref{am}) and Runge--Lenz vector (\ref{knut}) \cite{VV,
a7,a6,a4}.

We start with the observation that the angular momentum  and the
``relative electric charge'' (\ref{am}) are constructed with the
aid of the K vectors (\ref{kv}). The corresponding conserved
quantities in the spinning case are the followings:
\begin{gather*}
{\vec J}={\vec B} + {\vec j}, \qquad J_4 = B_4 + q
\end{gather*}
where we introduced the notation: ${\vec J} =(J_1, J_2, J_3)$,
${\vec B} = (B_1, B_2, B_3)$ and the spin corrections are
represented by the scalars $B_A$
\[
B_{A} = \tfrac i2 R_{A[\mu;\nu]}\psi^\mu \psi^\nu.
\]

Using equation (\ref{qf}) we can construct from the K-Y tensors
(\ref{fi}) and (\ref{fy}) the supercharges~$Q_i$ and $Q_Y$. The
supercharges $Q_i$ together with $Q_0$ from equation (\ref{Q0})
realize the $N=4$ supersymmetry algebra:
\[
\left\{ Q_A , Q_B \right\} = -2i\delta_{AB}H,\qquad A,B=0,\dots,3
\]
making manifest the link between the existence of the K-Y tensors
(\ref{fi}) and the hyper-K\" ahler geometry of the Taub-NUT
manifold. Moreover, the supercharges $Q_i$ transform as vectors at
spatial rotations
\[
\{Q_i ,J_j\}=\epsilon_{ijk} Q_k,\qquad i,j,k=1,2,3
\]
while $Q_Y$ and $Q_0$ behave as scalars.

To get the spin correction to the Runge--Lenz vector (\ref{knut})
it is necessary to investigate the generalized Killing
equations~(\ref{gke}) for $n=1$ with the S-K  tensor ${\vec
K}_{\mu\nu}$  in the right hand side. For an analytic expression
of the solution of this equation we shall use the decomposition
(\ref{kten}) of the S-K  tensor ${\vec K}_{\mu\nu}$ in terms of
K-Y tensors. Starting with this decomposition of the Runge--Lenz
vector ${\vec K}$ from the scalar case, it is possible to express
the corresponding conserved quantity ${\vec {\cal K}}$ in the
spinning case \cite{VV}:
\[
{\cal K}_i = 2m \left( -i\{ Q_Y , Q_i\} + \frac{1}{8 m^2} J_i J_4
\right).
\]

\section{Dirac equation on a curved background}

In what follows we shall consider the Dirac operator on a curved
background which has the form
\begin{equation}\label{DS}
D_s =  \gamma^\mu \hat\nabla_\mu.
\end{equation}
In this expression the Dirac matrices $\gamma_\mu$ are def\/ined
in local coordinates by the anticommutation relations
\[
\{ \gamma^\mu , \gamma^\nu \} = 2 g^{\mu\nu} I
\]
and $\hat\nabla_\mu$ denotes the canonical covariant derivative
for spinors. The essential properties of this covariant derivative
are summarized in the following equations
\begin{gather}
\hat\nabla_\mu \gamma^\mu = 0, \qquad \hat\nabla_{[\rho}
\hat\nabla_{\mu]} = \tfrac{1}{4} R_{\alpha\beta\rho\mu}
\gamma^\alpha \gamma^\beta,\nonumber
\end{gather}
where $R_{\alpha\beta\rho\mu}$ denotes the components of the
Riemann curvature tensor.

Carter and McLenaghan showed that in the theory of Dirac fermions
for any isometry with K vector $R_\mu$ there is an appropriate
operator \cite{CML}:
\[
X_k = -i \left( R^\mu \hat\nabla_\mu - \tfrac{1}{4} \gamma^\mu
\gamma^\nu R_{\mu;\nu}\right)
\]
which commutes with the {\it standard} Dirac operator (\ref{DS}).

Moreover each K-Y tensor $f_{\mu\nu}$ produces a {\it
non-standard} Dirac operator of the form
\begin{equation}\label{df}
D_f = -i\gamma^\mu \left(f_\mu ^{~\nu}\hat\nabla_\nu  -
\tfrac{1}{6}\gamma^\nu \gamma^\rho f_{\mu\nu;\rho}\right)
\end{equation}
which anticommutes with the standard Dirac operator $D_s$.

\section{Dirac equation in the Taub-NUT space}

When one uses Cartesian charts in the Taub-NUT geometry it is
useful to consider the local frames given by tetrad f\/ields $\hat
e(x)$, such that $g_{\mu\nu} = \delta_{\hat\alpha\hat\beta} \hat
e^{\hat\alpha}_\mu \hat e^{\hat\beta}_\nu$. The four Dirac
matrices $\hat\gamma^{\hat\alpha}$ that satisfy $\{
\hat\gamma^{\hat\alpha},\, \hat\gamma^{\hat\beta} \}
=2\delta^{\hat\alpha \hat\beta}$, can be taken as
\[
\hat\gamma^i = -i \left(
\begin{array}{cc}
0&\sigma_i\\
-\sigma_i&0
\end{array}\right),  \qquad
\hat\gamma^4 = \left(
\begin{array}{cc}
0&{\bf 1}_2\\
{\bf 1}_2&0
\end{array}\right),\qquad
\hat\gamma^5 = \hat\gamma^1\hat\gamma^2\hat\gamma^3\hat\gamma^4 =
\left(
\begin{array}{cc}
{\bf 1}_2&0\\
0&-{\bf 1}_2
\end{array}\right).
\]

In the Taub-NUT space the standard Dirac operator is
\[
{D}_{s}=\hat\gamma^{\hat\alpha}\hat\nabla_{\hat\alpha}
=i\sqrt{V}\vec{\hat\gamma}\cdot\vec{P} +
\frac{i}{\sqrt{V}}\hat\gamma^{4}P_{4} +\frac{i}{2}
V\sqrt{V}\hat\gamma^{4}\vec{\Sigma}^{*}\cdot\vec{B},
\]
where  $\hat\nabla_{\hat\alpha}$ are the components of the spin
covariant derivatives with local indices
\[
\hat\nabla_{i}=i\sqrt{V}P_{i}+\frac{i}{2}V\sqrt{V}\varepsilon_{ijk}
\Sigma_{j}^{*}B_{k},\qquad
\hat\nabla_{4}=\frac{i}{\sqrt{V}}P_{4}-\frac{i}{2}V\sqrt{V}
\vec{\Sigma}^{*}\cdot\vec{B}.
\]
These depend on the momentum operators
$P_{i}=-i(\partial_{i}-A_{i}\partial_{4})$, $P_{4}=-i\partial_{4}$
which obey the commutation rules $
[P_{i},P_{j}]=i\varepsilon_{ijk}B_{k}P_{4}$ and $[P_{i},P_{4}]=0$.
The spin matrices giving the spin connection are:{\samepage
\[
\Sigma_{i}^{*}=S_{i}+\tfrac{i}{2}\hat\gamma^{4}\hat\gamma^{i} ,
\qquad S_{i}=\tfrac{1}{2}\varepsilon_{ijk}S^{jk},
\]
where $S^{\hat\alpha \hat\beta}=-i [\hat\gamma^{\hat\alpha},\,
\hat\gamma^{\hat\beta}]/4$.}

In the above representation of the Dirac matrices, the Hamiltonian
operator of the {\em massless} Dirac f\/ield reads \cite{CV1,CV3}:
\[
H =\hat\gamma^5{D}_{s}=\left(
\begin{array}{cc}
0&V\pi^{*}\frac{\textstyle 1}{\textstyle \sqrt{V}}\\
\sqrt{V}\pi&0
\end{array}\right).
\]
This is expressed in terms of the operators
\[
\pi={\sigma}_{P}-\frac{iP_{4}}{V},\qquad
\pi^{*}={\sigma}_{P}+\frac{iP_{4}}{V},\qquad
\sigma_P=\vec{\sigma}\cdot\vec{P}
\]
and the Klein--Gordon operator has the form:
\[
\Delta= -\nabla_{\mu}g^{\mu\nu}\nabla_{\nu}=V\,\pi^{*}\pi=
V{\vec{P}\,}^{2}+\frac{1}{V}{P_{4}}^{2}.
\]

The conserved observables can be found among the operators which
commute or anticommute with $D_s$ and $\hat\gamma^5$
\cite{CV1,CV4}. That is the case of the total angular momentum
$\vec{J} =\vec{L} + \vec{S}$, where the orbital angular momentum
is
\[
\vec{L}\,=\,\vec{x}\times\vec{P}-4m\frac{\vec{x}}{r}P_{4} .
\]

Dirac-type operators are constructed from the K-Y tensors $f_i$
$(i=1,2,3)$ and $f_Y$ using equation (\ref{df}). In the quantum
Dirac theory these operators  replace the supercharges (\ref{qf})
from the pseudo-classical approach. Moreover we can give a {\it
physical} interpretation of the covariantly constant K-Y tensors
def\/ining the {\it spin-like} operators,
\begin{equation}\label{sl}
\Sigma_{i}=-\tfrac{i}{4}f^{i}_{\hat\alpha\hat\beta}
\gamma^{\hat\alpha}\gamma^{\hat\beta}=\left(
\begin{array}{cc}
\sigma_i&0\\
0&0
\end{array}\right),
\end{equation}
that have similar properties to those of the Pauli matrices. In
the pseudo-classical description of a Dirac particle
\cite{GRH,vH1}, the covariantly constant K-Y tensors correspond to
components of the spin which are separately conserved.

Here, since the Pauli matrices commute with the Klein--Gordon
operator, the spin-like operators (\ref{sl}) commute with $H^2$.
Remarkable the existence of the K-Y tensors allows one to
construct {\it Dirac-type} operators \cite{CML}
\[
Q_{i}=-if^{i}_{\,\hat\alpha\hat\beta}\gamma^{\hat\alpha}\hat\nabla
^{\hat\beta}=\{H,\,\Sigma_i\}
\]
which anticommute with $D_s$ and $\gamma^5$ and commute with $H$
\cite{CV3}. and obey the $N=4$ superalgebra, including $Q_0 = i
D_s = i \hat\gamma^5 H$:{\samepage
\[
\{Q_{A},\,Q_{B}\}=2\delta_{AB}H^2, \qquad A,B,\ldots=0,1,2,3
\]
linked to the hyper-K\" ahler geometry of the Taub-NUT space.}

Finally, using equation (\ref{df}), from the fourth K-Y tensor
$f_Y$ of the Taub-NUT space we can construct   the Dirac-type
operator \cite{CV1,CV7}
\[
Q_Y=\frac{r}{4m}\left\{H,\left(
\begin{array}{cc}
\sigma_{r}&0\\
0&-\sigma_{r} V^{-1}
\end{array}\right)
\right\} =i\frac{r}{4m}\left[Q_{0},\left(
\begin{array}{cc}
\sigma_{r}&0\\
0& \sigma_{r} V^{-1}
\end{array}\right)
\right].
\]

Let us observe that the Dirac-type operators $Q_A$ $(A=0,1,2,3)$
are characterized by the fact that their quantal anticommutator
close on the square of the Hamiltonian of the theory. No such
expectation applies to the non-standard, hidden Dirac operators
$Q^Y$ which close on a combination of dif\/ferent conserved
operators. The Taub-NUT space has a special geometry where the
covariantly constant K-Y tensors exist by virtue of the metric
being self-dual and the Dirac-type operators generated by them are
equivalent with the standard one. The fourth K-Y tensor $f^Y$
which is not covariantly constant exists by virtue of the metric
being of type $D$. The corresponding non-standard or hidden Dirac
operator does not close on $H$. It is associated with the hidden
symmetries of the space allowing the construction of the conserved
vector-operator analogous to the Runge--Lenz vector of the Kepler
problem.

The hidden symmetries of the Taub-NUT geometry are encapsulated in
the non-trivial S-K tensors $K_{i\mu\nu}$, $(i = 1,2,3)$. For the
Dirac theory the construction of the Runge--Lenz operator can be
done using products among the Dirac-type operators $Q_Y$ and
$Q_i$.

Let us def\/ine the operator \cite{CV1}:
\[
{N}_{i}=m\left\{ Q_{Y}, Q_{i}\right\}-J_{i}P_{4}.
\]

The components of the operator $\vec{N}$ commutes with $H$ and
satisfy the following commutation relations
\begin{gather*}
\left[{N}_{i},P_{4}\right]=0,\qquad
\left[{N}_{i},J_{j}\right]=i\varepsilon_{ijk}{N}_{k},\qquad
\left[{N}_{i},Q_{0}\right]=0,\\
\left[{N}_{i},Q_{j}\right]=i\varepsilon_{ijk}Q_{k}P_4 ,\qquad
\left[{N}_{i},{N}_{j}\right]= i\varepsilon_{ijk}J_{k}F^2
+\tfrac{i}{2}\varepsilon_{ijk} Q_{i}H\,\nonumber
\end{gather*}
where $F^2={P_4}^2-H^2$. In order to put the last commutator in a
form close to that from the scalar case \cite{GR,GM}, we can
redef\/ine the components of the Runge--Lenz operator, $\vec{{\cal
K}}$, as follows:
\[
{\cal K}_{i}={N}_{i}+ \tfrac{1}{2}H^{-1}(F-P_4) Q_i
\]
having the desired  commutation relation \cite{CV1}:
\[
\left[{\cal K}_{i},{\cal K}_{j}\right]=
i\varepsilon_{ijk}J_{k}F^2.
\]

\section{Gravitational anomalies}

For the classical motions, a  S-K tensor $K_{\mu\nu}$ generate a
quadratic constant of motion as in equation~(\ref{i1}). In the
case of the geodesic motion of classical scalar particles, the
fact that $K_{\mu\nu}$ is a S-K tensor satisfying (\ref{SK}),
assures the conservation of (\ref{i1}).

If we go from the classical motion to the hidden symmetries of a
quantized system, the corresponding quantum operator analog of the
quadratic function (\ref{i1}) is \cite{Carter}:
\begin{equation}\label{constq}
{\cal K} = D_\mu K^{\mu\nu} D_\nu,
\end{equation}
where $D_\mu$ is the covariant dif\/ferential operator on the
manifold with the metric $g_{\mu\nu}$. Working out the commutator
of (\ref{constq}) with the scalar Laplacian
\[
{\cal H} = D_\mu D^\mu
\]
we get from an explicit calculation gives \cite{Vogel}
\begin{gather*}
[D_\mu D^\mu, {\cal K}] = 2 K^{(\mu\nu;\lambda)} D_{\mu}D_\nu
D_{\lambda} + 3 K^{(\mu\nu;\lambda)}_{~~~~~~;\lambda} D_{\mu}
D_{\nu}\nonumber\\
\phantom{[D_\mu D^\mu, {\cal K}] =}{} + \left\{-\tfrac{4}{3}
K_\lambda^{~[\mu}R^{\nu]\lambda} + \tfrac {1}{2}g_{\lambda\sigma}(
K^{(\lambda\sigma;\mu);\nu} -
K^{(\lambda\sigma;\nu);\mu})+K^{(\lambda\mu;\nu)}_{~~~~~~;\lambda}
\right\}_{;\nu} D_\mu.
\end{gather*}
Note the very last terms are missing in the corresponding equation
in \cite{Carter}.

Concerning the hidden symmetry of the quantized system, the above
commutator does not vanishes on the strength of (\ref{SK}). If we
take $K$ to be a S-K tensor we are left with:
\begin{equation}\label{com2}
[{\cal H}, {\cal K}] = - \tfrac{4}{3}
\{K_\lambda^{~[\mu}R^{\nu]\lambda}\}_{;\nu} D_\mu
\end{equation}
which means that in general the quantum operator ${\cal K}$ does
not def\/ine a genuine quantum mechanical symmetry \cite{Car}. On
a generic curved spacetime there appears a {\it gravitational
quantum anomaly} proportional to a contraction of the S-K tensor
$K_{\mu\nu}$ with the Ricci tensor $R_{\mu\nu}$.

It is obvious that for a Ricci-f\/lat manifold this quantum
anomaly is absent. However, a more interesting situation is
represented by the manifolds in which the S-K tensor $K_{\mu\nu}$
can be written as a product of K-Y  tensors $f_{\mu\nu}$
\cite{CML}.

The integrability condition for any solution of (\ref{ky}),
written for K-Y tensors of valence $r=2$, is
\[
R_{\mu\nu[\sigma}^{~~~~\tau} f_{\rho]\tau} +
R_{\sigma\rho[\mu}^{~~~~\tau} f_{\nu]\tau} = 0.
\]
Now contracting this integrability condition on the Riemann tensor
for any solution of (\ref{ky}) we get
\begin{equation}\label{intcon2}
f^\rho_{~(\mu}R_{\nu)\rho} =0.
\end{equation}

Let us suppose that there exist a {\it square} of the S-K tensor
$K_{\mu\nu}$ of the form of a K-Y tensor $f_{\mu\nu}$  as in
equation~(\ref{i2}). In case this should happen, the S-K equation
(\ref{SK}) is automatically satisf\/ied and the integrability
condition (\ref{intcon2}) becomes
\[
K^\rho_{~[\mu}R_{\nu]\rho} =0.
\]

It is interesting to observe that in this latter equation an
antisymmetrization rather than symmetrization is involved this
time as compared to (\ref{intcon2}). But this relation implies the
vanishing of the commutator (\ref{com2}) for S-K tensors which
admit a decomposition in terms of K-Y tensors.

Using the S-K tensor components of the Runge--Lenz vector
(\ref{rl}) we can proceed to the evaluation of the quantum
gravitational anomaly for the generalized Taub-NUT metrics
\cite{IK1,IK2,IK3}. A direct evaluation \cite{tanug} shows that
the commutator (\ref{com2}) does not vanish.

To serve as a model for the evaluation of the commutator
(\ref{com2}) involving the components of the S-K tensors
corresponding to the Runge--Lenz vector (\ref{rl}), we limit
ourselves to give only the components of the third S-K
$k_3^{\mu\nu}$ tensor in spherical coordinates. Its non vanishing
components  are:
\begin{gather*}
k^{rr}_3 = -\frac{a r \cos\theta}{2(a+ b r)},\qquad
k^{r\theta}_3=k^{\theta r}_3=\frac{\sin\theta}{2},\qquad
k^{\theta\theta}_3 =\frac{(a + 2 b r) \cos\theta}{2r(a+ b r)},\\
k^{\varphi\varphi}_3= \frac{(a + 2 b r) \cot\theta\csc\theta}{
2r(a+ b r)},\qquad k^{\varphi\chi}_3=k^{\chi\varphi}_3=-\frac{(2a
+3br +
br\cos(2\theta)\csc^2\theta}{4r(a+br)},\nonumber\\
k^{\chi\chi}_3= \frac{(a - adr^2 + br(2 + cr)+ (a +
2br))\cot^2\theta)\cos\theta}{2r(a+br)}.\nonumber
\end{gather*}

Again, just to exemplify, we write down from the commutator
(\ref{com2}) only the function which multiplies the covariant
derivative $D_r$:
\begin{gather}
\frac{3r \cos\theta}{4(a+br)^3 (1 +cr +dr^2)^2}\{-2bd (2ad-bc) r^3
+
[3bd (2b-ac)-(ad+bc)(2ad-bc)]r^2\nonumber\\
\qquad{}+2(ad+bc)(2b-ac)r + a(2ad-bc)
+(b+ac)(2b-ac)\}\,.\label{angrav}
\end{gather}

Recall that the commutator (\ref{com2}) vanishes  for the standard
Euclidean Taub-NUT metric. It is easy to see that the above
expression (\ref{angrav}) vanishes for all $r$ if and only if the
constants $a$, $b$, $c$, $d$ are constrained by (\ref{standard}).

In conclusion the operators constructed from symmetric S-K
tensors are in general a source of gravitational anomalies for
scalar f\/ields. However, when the  S-K  tensor is of the form
(\ref{i2}), then the anomaly disappears owing to the existence of
the K-Y tensors.

\section{Index formulas and axial anomalies}

Atiyah, Patodi and Singer \cite{aps1} discovered an index formula
for f\/irst-order dif\/ferential operators on manifolds with
boundary with a non-local boundary condition. Their index formula
contains two terms, none of which is necessarily an integer,
namely a bulk term (the integral of a density in the interior of
the manifold) and a boundary term def\/ined in terms of the
spectrum of the boundary Dirac operator. Endless trouble is caused
in this theory by the condition that the metric and the operator
be of ``product type'' near the boundary.

In  \cite{tanug} we computed the index of the Dirac operator on
annular domains and on disk, with the non-local APS boundary
condition. For the generalized Taub-NUT metrics
\cite{IK1,IK2,IK3}, we found that the index is a number-theoretic
quantity which depends on the metrics. In particular, our formula
shows that the index vanishes on balls of suf\/f\/icient large
radius, but can be non-zero for some values of the parameters $c$,
$d$ and of the radius.

\begin{theorem}
If $c>-\frac{\sqrt{15d}}{2}$ then the extended Taub-NUT metric
does not contribute to the axial anomaly on any annular domain
(i.e., the index of the Dirac operator with APS boundary condition
vanishes).
\end{theorem}

\begin{proof}
The proof of this statement can be found in \cite{tanug}.
\end{proof}

The result is natural since  the index of an operator is unchanged
under continuous deformations of that operator. In our case this
would amount to a continuous change in the metric. The absence of
axial anomalies is due to the fact there exists an underlying
structure that does not depend on the metric. However for larger
deformations of the metric there could appear discontinuities in
the boundary conditions and therefore the index could present
jumps. Our formula for the index involves a computable
number-theoretic quantity depending on the parameters of the
metric.

We also examined the Dirac operator on the complete Euclidean
space with respect to this metric, acting in the Hilbert space of
square-integrable spinors. We found  that this operator is not
Fredholm, hence even the existence of a f\/inite index is not
granted.

We mentioned in \cite{tanug} some open problems in connection with
unbounded domains. The paper \cite{jpa} brings new results in this
direction. First we showed that the Dirac operator on $\mathbb
R^4$ with respect to the standard Taub-NUT metric does not have
$L^2$ harmonic spinors. This follows rather easily from the
Lichnerowicz formula, since the standard Taub-NUT metric has
vanishing scalar curvature. In particular, the index vanishes.

\begin{theorem}
For the standard Taub-NUT metric on $\mathbb R^4$ the Dirac
operator does not have $L^2$ solutions.
\end{theorem}
\begin{proof}
Recall that the standard Taub-NUT metric is hyper-K\"ahler, hence
its scalar curvature~$\kappa$ vanishes.

By the Lichnerowicz formula,
\[D^2=\nabla^*\nabla+\frac\kappa{4}=\nabla^*\nabla.\]

Let $\phi\in L^2$ be a solution of $D$ in the sense of
distributions. Then, again in distributions,
$\nabla^*\nabla\phi=0$.  The operator $\nabla^*\nabla$ is
essentially self-adjoint with domain
$\mathcal{C}^{\infty}_c(\mathbb R^4,\Sigma_4)$, which implies that
its kernel equals the kernel of $\nabla$. Hence $\nabla\phi=0$.
Now a parallel spinor has constant pointwise norm, hence it cannot
be in $L^2$ unless it is $0$, because the volume of the metric
${ds_K}^2$ is inf\/inite. Therefore $\phi=0$.
\end{proof}

\section{Concluding remarks}

In the pseudo-classical spinning particle models in curved spaces
from covariantly constant \mbox{K-Y} tensors $f_{\mu\nu}$ can be
constructed conserved quantities of the type $f_{\mu\nu} \psi^\mu
\psi^\nu$ depending on the Grassmann variables $\{\psi^\mu\}$
\cite{VV}. The Grassmann variables $\{\psi^\mu\}$ transform as a
tangent space vector and describe the spin of the particle. The
antisymmetric tensor $S^{\mu\nu} = - i \psi^\mu \psi^\nu$
generates the internal part of the local tangent space rotations.
For example, in the spinning Euclidean Taub-NUT space such
operators correspond to components of the spin which are
separately conserved~\cite{vH1}.

The construction of the new supersymmetries in the context of
pseudo-classical mechanics can be carried over straightforwardly
to the case of quantum mechanics by the usual replacement of phase
space coordinates by operators and Poisson--Dirac brackets by
anticommutators \cite{BeMa}. In terms of these operators the
supercharges are replaced by Dirac-type operators \cite{Ri}. In
both cases, the correspondence principle leads to equivalent
algebraic structures making obvious the relations between these
approaches \cite{vH1}.

In the study of  the Dirac equation in curved spaces, it has been
proved that the K-Y tensors play an essential role in the
construction of new Dirac-type operators. The Dirac-type operators
constructed with the aid of covariantly constant K-Y tensors are
equivalent with the standard Dirac operator \cite{Fort}. The
non-covariantly constant K-Y tensors generates  non-standard Dirac
operators which are not equivalent to the standard Dirac operator
and they are associated with the hidden symmetries of the space.

\looseness=1 The Taub-NUT space has a special geometry where the
covariantly constant K-Y tensors exist by virtue of the metric
being self-dual and the Dirac-type operators generated by them are
equivalent with the standard one. The fourth K-Y tensor $f^Y$
which is not covariantly constant exists by virtue of the metric
being of type $D$. The corresponding non-standard or hidden Dirac
operator does not close on $H$ and is not equivalent to the
Dirac-type operators. As it was mentioned, it is associated with
the hidden symmetries of the space allowing the construction of
the conserved vector-operator analogous to the Runge--Lenz vector
of the Kepler problem.

There is  a relationship between the absence of anomalies and the
existence of the K-Y tensors. For scalar f\/ields, the
decomposition (\ref{i2}) of S-K tensors in terms of K-Y tensors
guarantees the absence of gravitational anomalies. Otherwise
operators constructed from symmetric tensors are in general a
source of anomalies proportional to the Ricci tensors.

However for the axial anomaly the role of K-Y tensors is not so
obvious. The topological aspects are more important and the
absence of K-Y tensors does not imply the appearance of anomalies.

\appendix

\section{Euclidean Taub-NUT space}

Let us consider the Taub-NUT space and the chart with Cartesian
coordinates $x^\mu$ $(\mu, \nu =1,2,3,4)$ having the line element
\begin{equation}\label{metric}
ds^{2}=g_{\mu\nu}dx^{\mu}dx^{\nu}=f(r)(d\vec{x})^{2} +
\frac{g(r)}{16 m^2}(dx^{4}+ A_{i}dx^{i})^{2},
\end{equation}
where $\vec{x}$ denotes the three-vector $\vec{x} =
(r,\theta,\varphi)$,
$(d\vec{x})^{2}=(dx^{1})^{2}+(dx^{2})^{2}+(dx^{3})^{2}$ and
$\vec{A}$ is the gauge f\/ield of a monopole
\[
{\rm div}\vec{A}=0, \qquad \vec{B}={\rm rot}\,
\vec{A}=4m\frac{\vec{x}}{r^3}.
\]
The real number $m$ is the parameter of the theory which enter in
the form of the functions
\[
f(r) = g^{-1}(r) = V^{-1}(r) = \frac{4 m + r}{r}
\]
and the so called NUT singularity is absent if $x^4$ is periodic
with period $16 \pi m$. Sometimes it is convenient to make the
coordinate transformation $4 m (\chi + \varphi ) = - x^4$ with
$0\leq \chi < 4\pi$.

In the Taub-NUT geometry there are four K vectors \cite{GR,GM}
\begin{equation}\label{kv}
D_A=R_A^\mu\,\partial_\mu,\qquad A=1,2,3,4,
\end{equation}
where
\begin{gather}
D_1=-\sin\varphi\,\frac{\partial}{
\partial\theta}-\cos\varphi\cot\theta
\,\frac{\partial}{\partial\varphi}+\frac{\cos\varphi}{\sin\theta}
\frac{\partial}{\partial\chi},\nonumber\\
D_2=\cos\varphi\,\frac{\partial}{
\partial\theta}-\sin\varphi\cot\theta
\,\frac{\partial}{\partial\varphi}+\frac{\sin\varphi}{\sin\theta}
\frac{\partial}{
\partial\chi},\nonumber\\
D_3=\frac{\partial}{\partial\varphi},\qquad
D_4=\frac{\partial}{\partial\chi}.\label{kv4}
\end{gather}

$D_4$ which generates the $U(1)$ of $\chi$ translations, commutes
with the other K vectors. In turn the remaining three vectors,
corresponding to the invariance of the metric (\ref{metric}) under
spatial rotations ($A=1,2,3$), obey an $SU(2)$ algebra with
\[
[D_1, D_2]=-D_3, \qquad {\it etc}.
\]

In the bosonic case these invariances would correspond to the
conservation of angular momentum and the so called ``relative
electric charge'':
\begin{equation}\label{am}
\vec{j}=\vec{r}\times\vec{p}+q\frac{\vec{r}}{ r},\qquad q = g(r)
(\dot\theta + \cos\theta \dot\varphi),
\end{equation}
where $ \vec{p} = V^{-1}\dot{\vec{r}}$ is the mechanical momentum.

On the other hand, four K-Y tensors of valence 2 are known to
exist in the Taub-NUT geometry. The f\/irst three are covariantly
constant
\begin{gather}
f_i =8m(d\chi + \cos\theta d\varphi)\wedge dx_i
- \epsilon_{ijk}\left(1+\frac{4m}{r}\right) dx_j \wedge dx_k,\nonumber\\
D_\mu f^\nu_{i\lambda} =0, \qquad i,j,k=1,2,3.\label{fi}
\end{gather}
The $f^i$ def\/ine three anticommuting complex structures of the
Taub-NUT manifold, their components realizing the quaternion
algebra
\[
f^i  f^j + f^j f^i = - 2 \delta_{ij}, \qquad f^i  f^j - f^j f^i =
- 2 \varepsilon_{ijk} f^k.
\]

The existence of these K-Y tensors is linked to the hyper-K\"ahler
geometry of the manifold and shows directly the relation between
the geometry and the $N = 4$ supersymmetric extension of the
theory \cite{GRH,vH1}.

The fourth K-Y tensor is
\begin{equation}\label{fy}
f_Y = 8m(d\chi + \cos\theta  d\varphi)\wedge dr
+4r(r+2m)\left(1+\frac{r}{4m}\right)\sin\theta  d\theta \wedge
d\varphi
\end{equation}
having  a non-vanishing  covariant derivative
\[
{f_{Y}}_{r\theta;\varphi} =
2\left(1+\frac{r}{4m}\right)r\sin\theta.
\]

In  Taub-NUT space there is a conserved vector analogous to the
Runge-Lenz vector of the Kepler-type problem \cite{GR,FH,CFH}
\begin{equation}\label{knut}
\vec{K} = \tfrac{1}{2} \vec{K}_{\mu\nu}\dot x^\mu\dot x^\nu =
\vec{p}\times\vec{j} + \left(\frac{q^2}{4m}-
4mE\right)\frac{\vec{r}}{r},
\end{equation}
where the conserved energy is
\[
E = \tfrac{1}{2} g_{\mu\nu}\dot{x}^\mu \dot{x}^\nu.
\]
The components $K_{i\mu\nu}$ involved with the Runge--Lenz vector
(\ref{knut}) are St\"ackel--Killing tensors  satisfying the
equations
\[
K_{i(\mu\nu;\lambda)} = 0,\qquad K_{i\mu\nu} = K_{i\nu\mu}
\]
and they can be expressed as symmetrized products of the K-Y
tensors $f_i$, $f_Y$ and K vectors~$R_A$~\cite{VV}
\begin{equation}\label{kten}
K_{i\mu\nu} - \frac{1}{8m} (R_{4\mu} R_{i\nu} + R_{4\nu} R_{i\mu})
= m\big( f_{Y\mu\lambda} {{f_{i}}^\lambda}_\nu + f_{Y\nu\lambda}
{{f_{i}}^\lambda}_\mu \big).
\end{equation}

\section{Generalized Taub-NUT spaces}

In what follows we restrict ourselves to the  {\em generalized}
Taub-NUT manifolds whose metrics are def\/ined on ${\mathbb R}^4
\setminus \{0\}$ by the line element \cite{IK1,IK2,IK3}:
\begin{gather}
{ds_K}^2=g_{\mu\nu}(x)dx^{\mu}dx^{\nu}\nonumber\\
\phantom{{ds_K}^2}{}=f(r)(dr^2+r^2d\theta^2+r^2\sin^2\theta\,
d\varphi^2)
+g(r)(d\chi+\cos\theta\, d\varphi)^2, \label{mK}
\end{gather}
where the angle variables $(\theta,\varphi,\chi)$ parametrize the
sphere $S^3$ with $ 0\leq\theta<\pi$, $0\leq\varphi<2\pi$,
$0\leq\chi<4\pi$, while the functions
\[
f(r) = \frac{ a + b r}{r}  , \qquad g(r) = \frac{ a r + b r^2}{1 +
c r + d r^2}.
\]
depend on the arbitrary real constants $a$, $b$, $c$ and $d$.

Here it is worth pointing out that the above metrics are related
to the Berger family of metrics on 3-spheres \cite{hitch}.  These
are introduced starting with the Hopf f\/ibration $\pi_H$: $S^3\to
S^2$ that def\/ines the vertical subbundle $V \subset  TS^3$ and
its orthogonal complement $H \subset TS^3$ with respect to the
standard metric $g_{S^3}$ on $S^3$. Denoting with $g_H$ and $g_V$
the restriction of $g_{S^3}$ to the horizontal, respectively the
vertical bundle, one f\/inds that the corresponding line elements
are $d{s_H}^2=\frac{1}{4}{ds_2}^2$ and
${ds_V}^2=\frac{1}{4}({ds_3}^2-{ds_2}^2)$. For each constant
$\lambda >0$ the Berger metric on $S^3$ is def\/ined by the
formula
\[
g_{\lambda}=g_H+\lambda^2 g_V.
\]
This line element can be written in terms of the Berger metrics as
\[
{ds_K}^2=(ar+br^2)\left(\frac{dr^2}{r^2} +
4{ds_{\lambda(r)}}^2\right),
\]
where
${ds_{\lambda(r)}}^2=(g_{\lambda(r)})_{\mu\nu}dx^{\mu}dx^{\nu}$
and
\[
\lambda(r)=\frac{1}{\sqrt{1+cr+dr^2}}.
\]

If one takes the constants
\begin{equation}\label{standard}
c=\frac{2 b}{a}, \qquad d = \frac{b^2}{a^2}
\end{equation}
the generalized Taub-NUT metric becomes the original Euclidean
Taub-NUT metric up to a~constant factor.

By construction, the spaces with the metric (\ref{mK})  have four
K vectors (\ref{kv4}). The corresponding constants of motion in
generalized Taub-NUT backgrounds consist of a  conserved quantity
for the cyclic variable $\chi$
\[
q = g(r) (\dot\chi + \cos\theta \dot\varphi)
\]
and the angular momentum vector
\[
\vec{J}=\vec{x}\times\vec{p}+q\frac{\vec{x}}{r} , \qquad \vec{p} =
f(r)\dot{\vec{x}}.
\]

The remarkable result of Iwai and Katayama \cite{IK1,IK2,IK3}  is
that the generalized Taub-NUT space (\ref{mK})  admits a hidden
symmetry represented by a conserved vector, quadratic in
$4$-velocities, analogous to the Runge--Lenz vector of the
following form
\begin{equation}\label{rl}
\vec{K} = \vec{p} \times \vec{J} + \kappa \frac{\vec{x}}{r}.
\end{equation}
The constant $\kappa$ involved in the Runge--Lenz vector
(\ref{rl}) is $ \kappa = - a E + \frac{1}{2} c q^2 $ where the
conserved energy $E$ is
\[
E = \frac{\vec{p}^{~2}}{2 f(r)} + \frac{q^2}{2 g(r)} .
\]
The components $K_i=k^{\mu\nu}_i p_{\mu}p_{\nu}$ of the vector
$\vec{K}$ (\ref{rl}) involve three S-K tensors $k^{\mu\nu}_i$, $i
= 1,2,3$ satisfying (\ref{SK}).

\subsection*{Acknowledgements}

It is a pleasure to thank the organizers of the {\it
O'Raifeartaigh Symposium on Non-Perturbative and Symmetry Methods
in Field Theory} for the invitation to present this work. This
work is partially supported by a CEEX-MEC (Romania) Program.

\LastPageEnding
\end{document}